\documentclass{elsart}

\usepackage[ansinew]{inputenc} 
\usepackage[T1]{fontenc} 

\usepackage{graphics}

\usepackage{epsfig}

\usepackage{amssymb}
\usepackage{amsmath,cite}

\begin{document}

\begin{frontmatter}



\title{Nature and statistics of majority rankings in a dynamical model
of preference aggregation\thanksref{label1}}

 \thanks[label1]{We are indebted with G. Raffaelli for introducing us
 to this problem and for several important remarks and suggestions in
 the initial stages of this work.}


\author{G. L. Columbu$^1$, A. De Martino$^{1,2}$ and
A. Giansanti$^1$\corauthref{cor1}}

\address{$^1$Dipartimento di Fisica, Universit\`a di Roma `La Sapienza', Ple A. Moro 2, 00185 Roma, Italy\\
$^2$INFM-CNR (ISC), Roma, Italy}

\corauth[cor1]{Corresponding author. E-mail: andrea.giansanti@roma1.infn.it}

\begin{abstract}
We present numerical results on a complex dynamical model for the
aggregation of many individual rankings of $S$ alternatives by the
pairwise majority rule under a deliberative scenario. Agents are
assumed to interact when the Kemeny distance between their rankings is
smaller than a range $R$. The main object of interest is the
probability that the aggregate (social) ranking is transitive as a
function of the interaction range. This quantity is known to decay
fast as $S$ increases in the non-interacting case. Here we find that
when $S>4$ such a probability attains a sharp maximum when the
interaction range is sufficiently large, in which case it
significantly exceeds the corresponding value for a non-interacting
system. Furthermore, the situation improves upon increasing $S$. A
possible microscopic mechanism leading to this counterintuitive result
is proposed and investigated.
\end{abstract}

\begin{keyword}
social choice, Condorcet paradox, pairwise majority rule
\PACS 89.65.-s, 05.10.-a, 02.50.Ey
\end{keyword}
\end{frontmatter}

\section{Introduction}

The aggregation of many individual preferences into a single,
``social'' preference is a long-studied problem in mathematical social
sciences which has more recently also been considered from a
statistical mechanics viewpoint. Here by ``preference'' we mean simply
a ranking of say $S$ objects, i.e. a complete ordering e.g. from the
favorite downwards. A minimal consistency requirement for preferences
is that they should be transitive, i.e. if $A$ is preferred to $B$ and
$B$ to $C$, then $A$ should be preferred to $C$. It is however well
known that as soon as $S$ exceeds 2 the construction of the aggregate
preference runs into the Condorcet problem \cite{condo,dl,das}:
starting from transitive individual preferences, the social preference
may turn out to be intransitive, i.e. contain a cycle of the type
$A\succ B\succ C\succ A$ (meaning A preferred to B etc.). The
probability with which an intransitive ranking emerges from random
individual preferences (`impartial culture assumption' \cite{g83}) has
been studied in the mathematical economics literature in the past
\cite{nw,gf,ggg}. Remarkably, this bottleneck is present to different
degrees for all aggregation methods one considers and can be removed
only at the cost of loosening some of the requirements that a social
choice should satisfy\footnote{This is for example the case of
plurality voting, which avoids the Condorcet problem but is vulnerable
to tactical voting (see also \cite{das}).}.

Among the different aggregation methods, the pairwise-majority rule
(PMR) has been the most studied as it appears to be more robust to the
above mentioned requirements. According to PMR, the social preference
corresponds to that derived by comparing objects pairwise via a simple
majority rule. Recently it has been possible to quantify the extent to
which PMR is effective as a social choice rule by computing the
probability that the aggregate of $N\gg 1$ random preferences is
transitive within a statistical mechanics framework \cite{rm}. It is
similarly important to understand how the situation changes if voters
interact before casting their ballot. In \cite{rm}, a random-field
type of interaction with conformism has been considered, and it has
been shown, among other results, that an interacting population may
reach consensus on one of $S$ issues more easily the larger is $S$.  A
scenario that has received much attention recently is the so-called
deliberative approach \cite{list}, where voters discuss the
alternatives before the vote and eventually change their
preferences. The idea is that discussion should lead to preference
harmonization and structuring and thus drastically reduce the
probability of an intransitive social choice.

In this paper we study the deliberative scenario in a dynamical model
of $N$ interacting agents or voters, whose preferences are initialized
to transitive random ones to simulate the original diversity of
opinions. The key ingredient of the model is that an agent interacts
only with agents whose preferences are sufficiently close to his. In
such ``neighborhoods'', by conformity he aligns to the PMR-aggregate
preference (if transitive). This generalizes a classical social
interaction mechanism investigated previously in
e.g. \cite{galam,wd}. As a measure of similarity between orderings we
use the Kemeny distance \cite{ks,sm}, widely used in the social
sciences (it is related to their Hamming distance). This choice is
arbitrary and it is likely that distances more sensitive to the
position in the orderings may give different results. A social
preference is then formed via PMR after every agent has updated his
preference.

We are interested in studying the behavior of a specific macroscopic
observable, namely the probability of a transitive aggregate ranking
as a function of time (number of system-wide updates) and of the range
of interaction (the maximum distance within which agents
interact). Our main result is that there is an optimal interaction
range (or more properly a window of ranges) for which the Condorcet
problem is much less likely to occur than in the non-interacting case,
and that the frustration decreases as a function of both time and
$S$. The model will be fully defined in the following section. Its
complexity has prevented analytical approaches on our side. Our
results are thus obtained by means of numerical simulations.

\section{Definition of the model}

Consider $N$ agents (labeled by $i,j,\ldots$) each of whom ranks $S$
alternatives ${a_1,a_2,...,a_S}$. A ranking is a complete transitive
ordering such as $a_1\succ a_2\succ a_3\succ ...\succ a_S$. At time
zero, every agent possesses a transitive preference ranking selected
randomly with uniform probability among the $S!$ possible ones. Every
ranking can be split uniquely into pair-wise comparisons of $S(S-1)/2$
pairs of distinct alternatives. We label pairs as
$\alpha,\beta,\ldots$ and denote by $Q_i^{(\alpha)}$ agent $i$'s
preferences on pair
$\alpha=(a_{\alpha_1},a_{\alpha_2})$. Specifically, $Q_i^{(\alpha)}=1$
if $a_{\alpha_1}\succ a_{\alpha_2}$ and $Q_i^{(\alpha)}=-1$ otherwise
(ties are excluded). The dissimilarity between the rankings of agents
$i$ and $j$ is given by their Kemeny distance
\begin{equation} 
K(i,j)=\dfrac{2}{S(S-1)}\sum_{\alpha}\left(1-\delta_{Q_i^{(\alpha)},Q_j^{(\alpha)}}\right),
\end{equation}
where the sum runs over all $S(S-1)/2$ pairs. $K(i,j)$ measures simply
the normalized number of pairs that are ranked differently by agents
$i$ and $j$, irrespective of the position in the alignment, or in
other words it is the number of adjacent pairwise switches needed to
convert one preference order into the other.

We introduce an interaction range $0\leq R\leq 1$ and define the
neighborhood of $i$ as the set of agents whose rankings have a Kemeny
distance of at most $R$ from his:
\begin{equation}
V(i)=\{j~\text{such that}~K(i,j)\leq R\}
\end{equation}
At every time step, an agent is selected randomly and interaction
takes place. Specifically, the agent changes his ranking to that
derived from a PMR among agents in his neighborhood if the latter is
transitive, otherwise he keeps his preference unchanged. In subsequent
interactions the agents enters with his new ranking. After a
system-wide update (sweep) is performed a global vote by PMR takes
place, the social ranking is computed and agents move into the next
round. We describe details of the PMR procedure below (see also
\cite{rm}). We notice that as $R$ decreases from 1 to 0 the
interaction becomes more and more of a local nature. However it is to
be expected that the number of surviving rankings decreases in time as
the local interaction mimics conformist behavior on the side of
agents. Furthermore, neighborhoods evolve in time (also inside a
single system-wide update).

The basic function which is numerically evaluated in this study is the
probability that PMR yields a social transitive ranking, denoted by
$P(S)$. It is evaluated by counting the number of times a collective
transitive order is obtained through PMR, out of a large number of
samples. We monitor the evolution of $P(S)$ in time specifically
varying the number $U$ of sweeps. In absence of interaction, $P(S)$
decays as $S$ increases, though less fast than the na\"\i ve guess
$S!/2^{S(S-1)/2}$ corresponding to the ratio of the number of
transitive rankings to the total number of binary vectors encoding
different rankings of $S$ objects \cite{ggg,rm}.

Coming to the details of PMR voting, for each pair $\alpha$, let
\begin{equation}
M^{(\alpha)} = \Theta\left(\sum_{i=1}^{N} Q_i^{(\alpha)}\right),
\end{equation}
where $\Theta(x)$ is Heaviside function. Clearly, $M^{(\alpha)}=1$ if
the majority of agents prefers $a_{\alpha_1}$ over $a_{\alpha_2}$,
whereas $M^{(\alpha)}=0$ if the majority ranks $a_{\alpha_2}$ over
$a_{\alpha_1}$. The aggregate order (either local or global) emerges
from separate majority votes over all pairs, that is from a
computation of the $M^{(\alpha)}$, for all $\alpha$. There is a simple
method to check whether a ranking defined through the different
$M^{(\alpha)}$'s is transitive. Indeed each $M^{(\alpha)}$ can be seen
as the element of a $S\times S$ matrix since
$\alpha=(\alpha_1,\alpha_2)$. Let us then write explicitly
$M^{(\alpha)}$ as $M^{(\alpha_1,\alpha_2)}$ and let
\begin{equation} 
C_{\alpha_1}=\sum_{\alpha_2=1}^{S} M^{(\alpha_1,\alpha_2)}
\end{equation}
One easily understands (e.g. by induction starting from small $N$ and
$S$) that if the PMR ranking is transitive, then the $S$-vector
$\vec{C}$ will contain once and only once each of the integers
$0,1,2,...,S-1$ as elements.

\section{Results}

\begin{figure}
 \centering
 \includegraphics*[width=10cm]{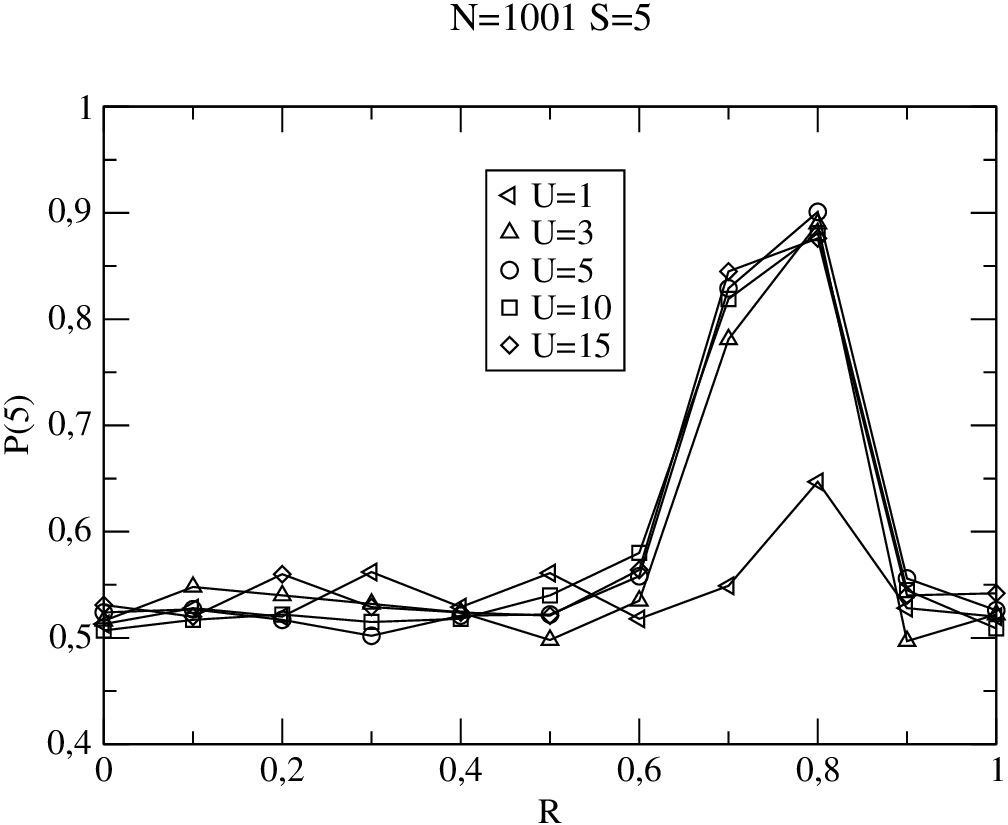}
\caption{\label{figure3} $P(S)$ as a function of the interaction range
$R$ and of the number of sweeps $U$. Parameters: $N=1001$, $S=5$;
average over $1000$ samples.}
\end{figure}

In Fig. \ref{figure3} we display the time evolution (in units of
sweeps) of $P(5)$ as a function of $R$ for a system of size $N=1001$
(the choice of $5$ alternatives is here only a matter of convenience;
qualitatively identical results occur for larger values of $S$).
It is evident that after just a cycles the dependence of $P(5)$ on the
radius of interaction becomes stable. It is clear, also from this
graph, the onset of a peculiar regime in a window of interaction radii
comprised between $R=0.6$ and $R=0.9$, where the probability of
getting a transitive ranking is considerably enhanced with respect to
the non interacting case. It is also worth noting that, until $R$ is
sufficiently small, $P(5)$ is almost constant and coincident with the
corresponding probability in the non interacting case; then it
abruptly increases and reaches the value of $0.9$ at $R=0.8$. For
larger ranges, $P(5)$ decreases again to the non-interacting case.

\begin{figure}
 \centering
 \includegraphics*[width=10cm]{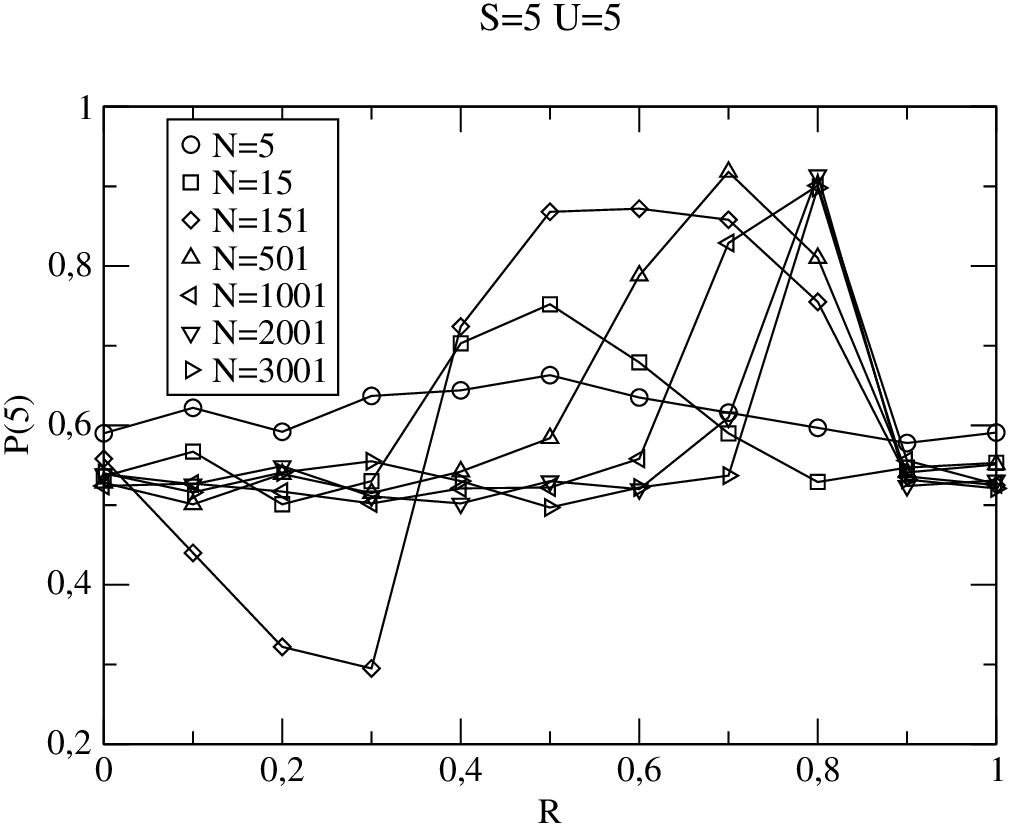}
\caption{\label{figure2}$P(5)$ as a function of the radius of
interaction $R$ and the number of agents $N$ after $U=5$
sweeps. Average over $1000$ samples. For $R=0$ and $R=1$ one recovers
the value observed for the non interacting case.}
\end{figure}

In order to characterize the scaling of the distribution with
increasing system size, in Fig. \ref{figure2} we show the dependence
of $P(5)$ on the number of agents $N$ and on the radius of interaction
$R$ (the asymptotic states is reached in all cases shown). The
reference value for a non-interacting system is $P(5)\simeq 0.54$.

\begin{figure} 
\centering \includegraphics*[width=10cm]{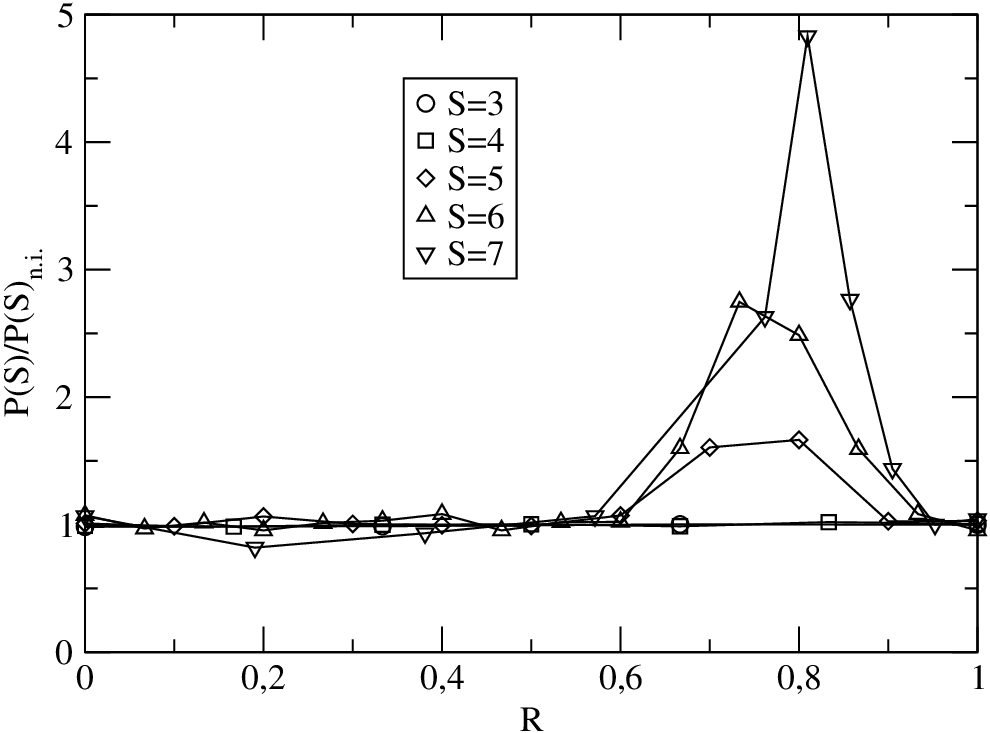}
\caption{\label{figure4}$P(S)$ normalized by the corresponding
$P(S)_{n.i.}$ for non interacting agents as a function of
$R$. Parameters are the same as in the previous figure, except for
$S=7$, for which $N=6001$ and the average is performed over 500
samples. Here, $U=15$.}
\end{figure}

Remarkably, the existence of an optimal interaction range is
reinforced by increasing the number of alternatives, as shown in
Fig. \ref{figure4}.
Note that for $S=4$ the probability of getting a transitive majority
ranking is still that of the non-interacting case. For $S>5$ this
probability increases, with a gradual gain which abruptly falls down
for larger $R$.

\begin{figure}
 \centering \includegraphics*[width=10cm]{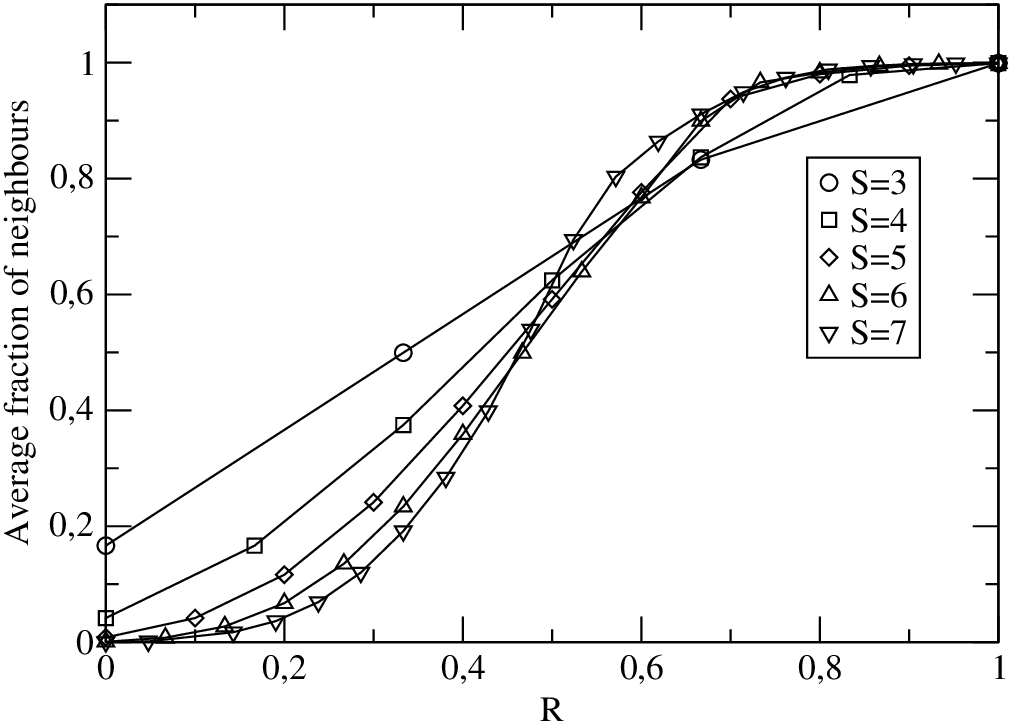}
\caption{\label{figure5}Average fraction of neighbors as a function
of $R$ and $S$. Parameters are the same as in the previous figure.}
\end{figure}

To shed light on these observations, we report in Fig. \ref{figure5},
for different $S$, the $R$-dependence of the average fraction of
neighbors, i.e. the average number of agents with preference orders
whose Kemeny distance does not exceed $R$.
This parameter is computed, after $U$ sweeps, by recording the number
of neighbors of each agent in a sample and averaging over samples. By
increasing $S$ there is a marked tendency of the curve to acquire a
sigmoidal shape. The average fraction of neighbors increases
gradually, with a maximum rate at $R\simeq 0.5$, and saturates to $1$
for large enough $R$. This result indicates that there is a
cooperative transition between a local regime and a global,
effectively long range, regime. In the former, each agent interacts
and confronts its preference order only with a small mass of the other
agents; in the latter, each agent's opinion is influenced by the
opinions of most of the others. Note that in the cases $S=3$ and $S=4$
the $R$ dependence is monotonously increasing, but no sign of
cooperativity is present yet, in accordance with the result of the
previous figure.

\begin{figure}
\centering \includegraphics*[width=10cm]{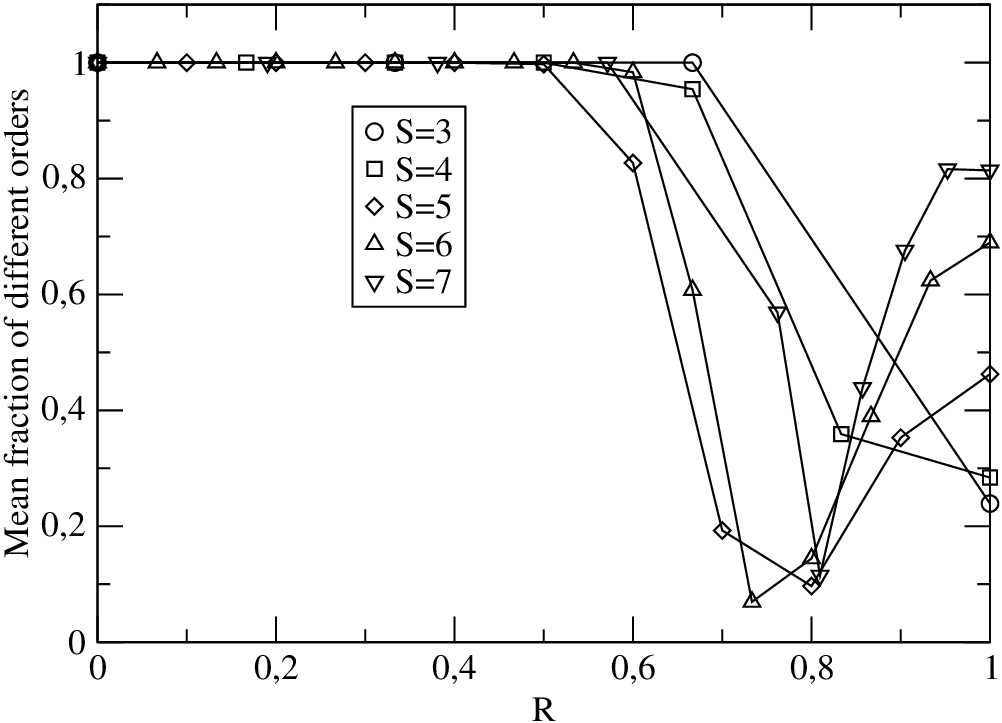}
\caption{\label{figure6}Mean fraction of different transitive
preference orders surviving after $U=15$ sweeps as a function of
$R$. Parameters are the same as in the previous figure.}
\end{figure}

\begin{figure}
\centering \includegraphics*[width=10cm]{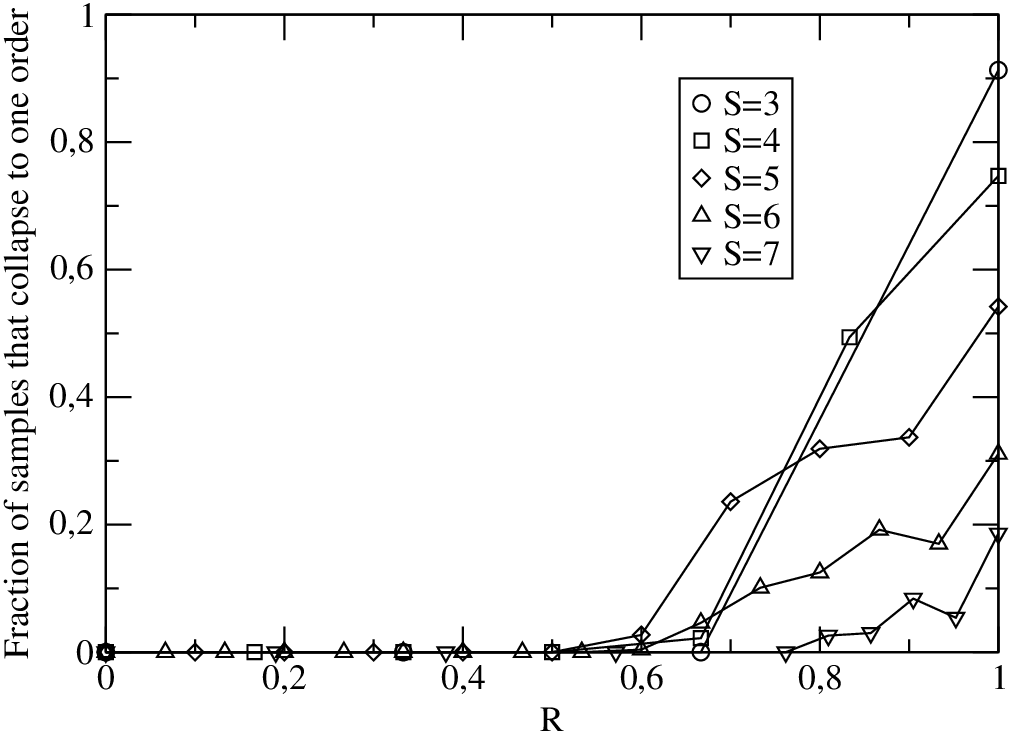}
\caption{\label{figure7}Fraction of samples in which only one ranking
survives after $U=15$ sweeps as a function of $R$. Parameters are the
same as in the previous figure.}
\end{figure}

To investigate further this transition, we consider in
Fig. \ref{figure6} the $R$ dependence of the mean fraction of
different transitive preference orders present in a population of
agents after a long update cycle.
Remember that initially, before the update cycle starts, transitive
preference orders are randomly distributed among the agents. Below
$R\simeq 0.6$ the interaction preserves all the possible diversity of
preference orders, while above the cooperative transition has taken
place and there is a reduction of diversity. For $S>5$ at $R=0.8$ only
about ten percent of the possible different transitive rankings are
still present among the agents. Note that for sufficiently large $R$
one observes an increase of different transitive preference order, as
is to be expected in view of the fact that the interaction mechanism
in systems with very large and very small $R$ is qualitatively
similar. This phenomenon should be confronted with the result in
Fig. \ref{figure7}, where the $R$-dependence of the fraction of
samples with only one surviving preference order.
Note, that at $R=1$ this fraction coincides, for different $S$, with
the corresponding $P(S)$ values of the non interacting case. Figures
\ref{figure6} and \ref{figure7} show that update interaction cycles
induce, if the cooperative transition is established, a reduction of
the different transitive preference orders among which to choose in
the PMR. Moreover, for large enough interaction radii, when
practically each agents interacts with all the others, there is a
marked tendency towards the emergence of a dominant transitive
preference order. As is to be expected, this tendency is contrasted by
the increase in $S$, the number of alternatives.

\begin{figure}
\centering \includegraphics*[width=10 cm]{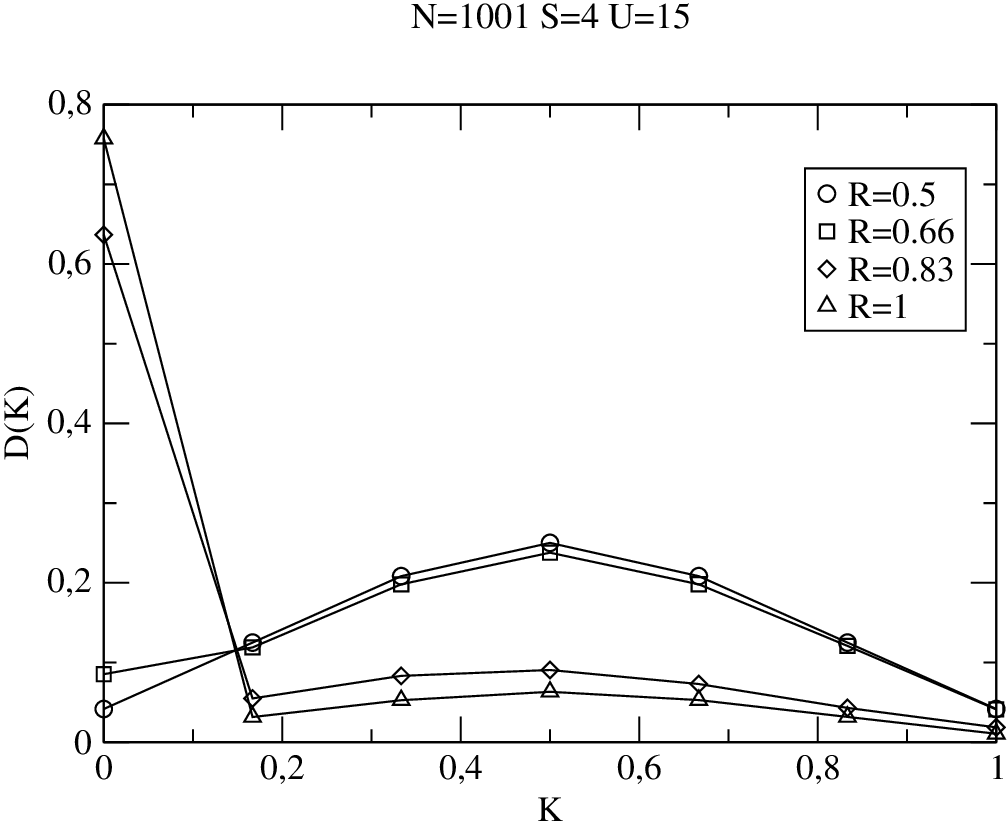}
\caption{\label{figure8}Pair distribution function $D(K)$ of the
Kemeny distances between the preference orders of pairs of different
agents, after the interaction process. $S=4$ $U=15$; $N=1001$ and
$1000$ samples.}
\end{figure}

\begin{figure}
\centering \includegraphics*[width=10 cm]{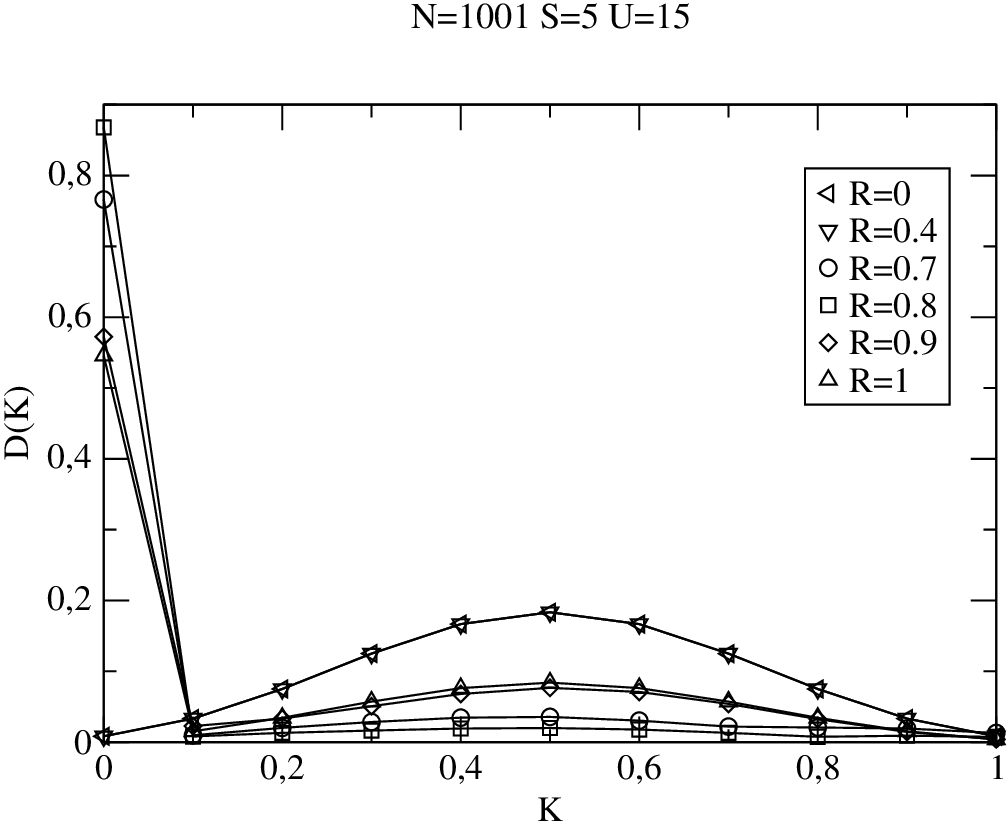}
\caption{\label{figure9}Pair distribution function $D(K)$ of the
Kemeny distances $K$ between the preference orders of pairs of
different agents, after the interaction process. $S=5$ $U=15$;
$N=1001$ and $1000$ samples.}
\end{figure}

In Fig. \ref{figure8} and \ref{figure9} we show the pair distribution
function of the Kemeny distances among individual rankings, as a
function of $R$. We consider in these figures the discriminating
cases: $S=4$ and $S=5$.
In both cases, if $R$ is less than $0.5$, the pair distribution has a
maximum for $K=0.5$ (corresponding to $S(S-1)/4$ adjacent pairwise
switches needed to convert one ranking into the other). This is
essentially the pair distribution function corresponding to the
initial random assignment of preference orders. As an effect of the
interaction it is seen that, when the cooperative transition is
completed, i.e. for large enough $R$, a peak at $K=0$ emerges,
corresponding to the formation of a big clique of agents with the same
transitive preference order. This points to an interesting herding
effect, whose details deserve further investigation.

\section{Discussion and Conclusions}

We have studied the emergence of the Condorcet problem in a
deliberative multi-agent scenario. We have studied in particular if and
how the frustration expressed by Condorcet's paradox can be mitigated
by interaction among the agents, before the global PMR voting takes
place. We have introduced an interaction scheme based on local PMR
voting among ``neighboring'' agents, whose preference orders are
close, to mimic conformist behavior. Our results point to the
existence of two regimes controlled by the interaction range $R$, with
a crossover from one to the other at intermediate $R$. For low $R$,
the probability of getting a transitive outcome is unaffected (with
respect to the non-interacting case), whereas for sufficiently large
$R$ a marked enhancement in $P(S)$ is observed, which increases with
$S$, the number of alternatives to be ranked. A herding phenomenon is
furthermore observed which reduces the repertoire of different
surviving rankings. So, if the radius of interaction is too large it
is difficult to have a transitive PMR outcome, but in the case it is
reached that happens because all the agents practically vote in the
same way.


These results support the claims that deliberative systems reduce the
chance for the formation of cycles in the social choice, provided the
interaction range lies in the optimal window. It would be important to
get a deeper insight on the microscopic details of this model. In
particular the computation of two point correlations, like the
probability that after an interaction cycle two agents taken at random
belong to the same neighborhood, or the probability that after the
interaction two initially separated neighborhoods become
overlapping. The further investigation of the clustering dynamics
(including coalescence of neighborhoods) could provide important
insight also in directions different than statistical mechanics
\cite{mb}.





\end{document}